\documentclass[
 reprint,
 superscriptaddress,
 amsmath,amssymb,
 aps,
pra,
showkeys
]{revtex4-2}

\usepackage{soul}
\usepackage{float}
\usepackage{multibib}
\usepackage{hyperref}
\usepackage{xcolor}
\hypersetup{
    colorlinks,
    linkcolor={blue!50!blue},
    citecolor={blue!50!blue},
    urlcolor={blue!80!black}
}
\usepackage{siunitx}
\usepackage{import}
\usepackage{placeins}
\usepackage{xcolor}
\usepackage{lmodern}
\usepackage[braket, qm]{qcircuit}
\usepackage{graphicx}
\usepackage{dcolumn}
\usepackage{bm}
\usepackage{comment}
\usepackage[english]{babel}
\usepackage{blindtext}
\usepackage{physics}  
\usepackage{multirow}
\usepackage{array}
\usepackage{enumerate}

\usepackage{ulem}
\definecolor{myblue}{rgb}{0.0, 0.75, 1.0}
\definecolor{lightpink}{rgb}{0.9, 0.4, 0.38}

\begin{document}
\title{The hBN defects database: a theoretical compilation of color centers in hexagonal boron nitride}  

\author{Chanaprom Cholsuk}
\email{chanaprom.cholsuk@tum.de}
\affiliation{Department of Computer Engineering, TUM School of Computation, Information and Technology, Technical University of Munich, 80333 Munich, Germany}
\affiliation{Munich Center for Quantum Science and Technology (MCQST), 80799 Munich, Germany}
\affiliation{Abbe Center of Photonics, Institute of Applied Physics, Friedrich Schiller University Jena, 07745 Jena, Germany}

\author{Ashkan Zand}
\affiliation{Department of Computer Engineering, TUM School of Computation, Information and Technology, Technical University of Munich, 80333 Munich, Germany}
\affiliation{Munich Center for Quantum Science and Technology (MCQST), 80799 Munich, Germany}
\affiliation{Abbe Center of Photonics, Institute of Applied Physics, Friedrich Schiller University Jena, 07745 Jena, Germany}

\author{Asl{\i} \surname{\c{C}akan}}
\affiliation{Department of Computer Engineering, TUM School of Computation, Information and Technology, Technical University of Munich, 80333 Munich, Germany}
\affiliation{Munich Center for Quantum Science and Technology (MCQST), 80799 Munich, Germany}

\author{Tobias Vogl}%
\email{tobias.vogl@tum.de}
\affiliation{Department of Computer Engineering, TUM School of Computation, Information and Technology, Technical University of Munich, 80333 Munich, Germany}
\affiliation{Munich Center for Quantum Science and Technology (MCQST), 80799 Munich, Germany}
\affiliation{Abbe Center of Photonics, Institute of Applied Physics, Friedrich Schiller University Jena, 07745 Jena, Germany}

\date{\today}

\begin{abstract}
Color centers in hexagonal boron nitride (hBN) have become an intensively researched system due to their potential applications in quantum technologies. There has been a large variety of defects being fabricated, yet, for many of them, the atomic origin remains unclear. The direct imaging of the defect is technically very challenging, in particular since, in a diffraction-limited spot, there are many defects and then one has to identify the one that is optically active. Another approach is to compare the photophysical properties with theoretical simulations and identify which defect has a matching signature. It has been shown that a single property for this is insufficient and causes misassignments. Here, we publish a density functional theory (DFT)-based searchable online database covering the electronic structure of hBN defects (257 triplet and 211 singlet configurations), as well as their photophysical fingerprint (excited state lifetime, quantum efficiency, transition dipole moment and orientation, polarization visibility, and many more). All data is open-source and publicly accessible at \url{https://h-bn.info} and can be downloaded. It is possible to enter the experimentally observed defect signature and the database will output possible candidates which can be narrowed down by entering as many observed properties as possible. The database will be continuously updated with more defects and new photophysical properties (which can also be specifically requested by any users). The database therefore allows one to reliably identify defects but also investigate which defects might be promising for magnetic field sensing or quantum memory applications.
\end{abstract}


\maketitle

\section{Introduction}
For single photon emission in two-dimensional (2D) materials, a two-level system is required, which either originates from a point-like defect inducing defect states into the band gap, or from local band bending (localized excitons). Such 2D materials are, for instance, transition metal dichalcogenides (TMDs) \cite{Dang2020,Li2022,Parto2021,Lee2022,Gupta2019} and hexagonal boron nitride (hBN) \cite{Tran2016,Vogl2019-ACS,Kumar2023}. Quantum emission from hBN is particularly appealing, as it covers a photon emission range from UV \cite{doi:10.1021/acs.nanolett.6b01368} to near-infrared \cite{10.1063/5.0008242} and is therefore potentially compatible with a variety of applications. Moreover, single photon emission from hBN is bright, can be controlled using polarization \cite{Kumar2024,Jungwirth2017}, and tailored using fine-tuning techniques \cite{Lyu2022,Li2020,Cholsuk2022}. The demonstrated applications range from quantum sensing (magnetic fields \cite{Stern2022,Mu2022,Gottscholl2020}, strain \cite{Grosso2017}, pressure \cite{Gottscholl2021}, temperature \cite{doi:10.1021/acsami.0c05735}), space instrumentation \cite{Vogl2019space,https://doi.org/10.1002/qute.202300343}, quantum cryptography \cite{https://doi.org/10.1002/qute.202200059}, quantum computing \cite{Conlon2023}, quantum random number generation \cite{10.1063/5.0074946}, and fundamental quantum physics experiments \cite{PhysRevResearch.3.013296}.\\
\indent The positions of defect states in host materials strongly depend on the type of defects; some are localized at shallow levels, while others are localized at deep levels \cite{Xiong_2024,Cholsuk2022,Tawfik2017,Sajid2018}. The electronic structure of any defect can be calculated with density functional theory (DFT). Although this calculation can reasonably predict the electronic structures, assigning such microscopic origins remains challenging due to excessive defect choices and similar properties of different defects. The former results from the intrinsic wide-band gap allowing hBN to host many defects \cite{Cholsuk2022,Tan2022,Sajid2018}, while the latter is observed in experiments, e.g., there are many so-called 2 eV emitters which might not have the same atomic origin and vary in other photophysical properties \cite{Vogl2019,Kumar2023,Sajid2020,Li2022-2eV}. Therefore, accurate defect identification requires not only a comprehensive dataset to cover all possible defects but also sophisticated theoretical approaches to accurately capture the electronic properties \cite{Cholsuk2023}. \\
\begin{figure*}[ht]
    \centering
    \includegraphics[width = 1\textwidth]{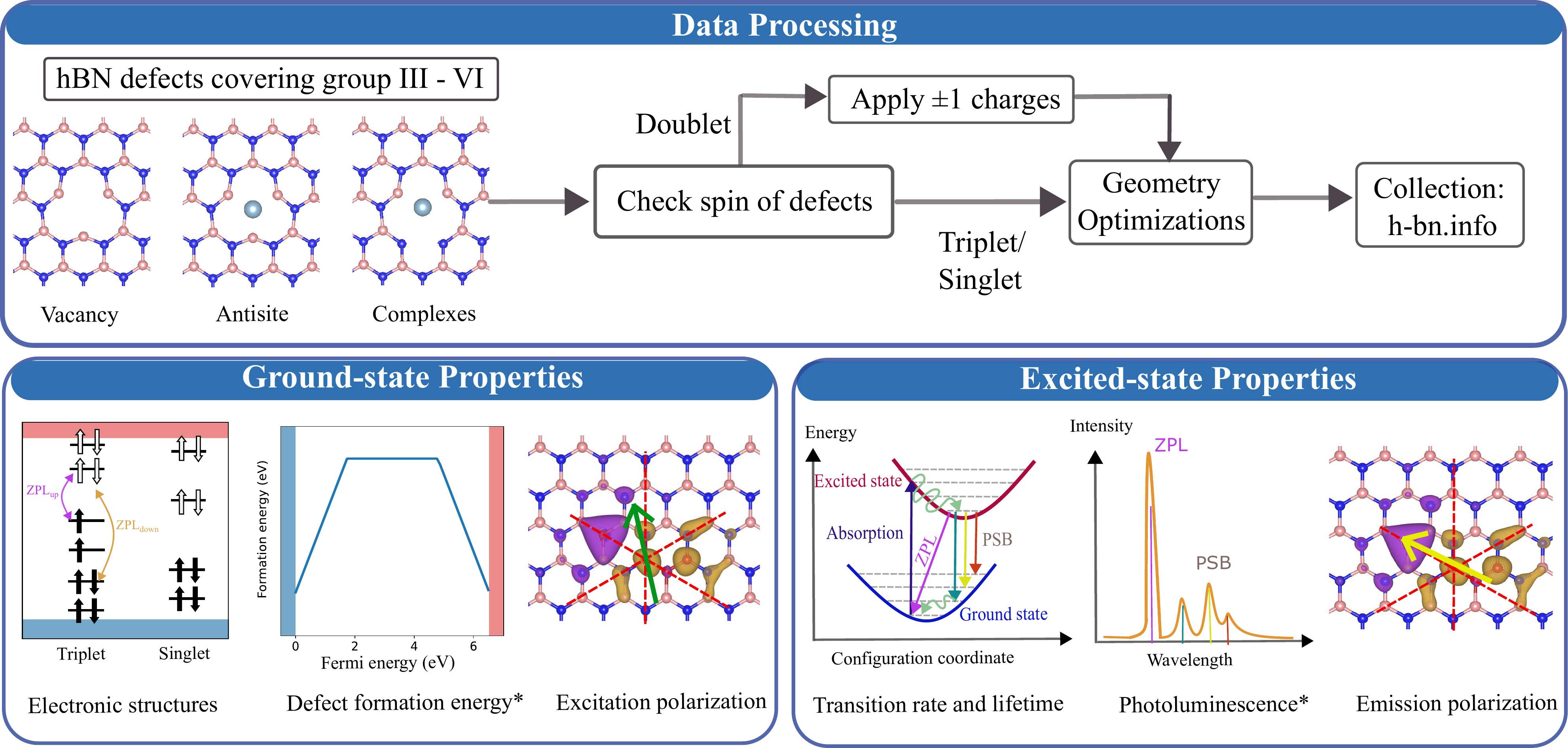}
    \caption{Flowchart of the DFT database acquisition. A DFT calculation was initially performed to investigate hBN defects. After that, ground and excited photophysical properties over 257 triplet transitions were extracted. Note that the symbol * denotes properties scheduled for future updates within the database (and are only currently available for a subset of defects). This database is openly accessible via \url{https://h-bn.info}. Some subfigures are adapted from Ref.~\cite{Cholsuk2024} available under a CC-BY 4.0 license. Copyright 2024 Cholsuk \textit{et al}. Some subfigures are adapted with permission from Ref.~\cite{Cholsuk2023}. Copyright 2023 American Chemical Society.}
    \label{fig:workflow}
\end{figure*}
\indent Recently, numerous databases have vastly been developed to collect properties of a large number of defects in 2D materials. These include machine-learning-based databases \cite{Shenoy2020,Chen2019,Arun2020} and DFT-based databases \cite{2DMatPedia,MC2D,MC2D-2,C2DB,C2DB-update,QPOD, IMP2D, Ali2023}. The former can handle a large number of datasets but the prediction of promising candidates still requires a DFT calculation for confirmation. In general, a machine-learned database is sufficient for predicting promising new defects that one could fabricate but for defect identification one needs the most accurate data. This can be offered by DFT calculations: the available databases contain structural, electronic, and energetic properties \cite{2DMatPedia,MC2D,MC2D-2,C2DB,C2DB-update}, Raman spectra \cite{Taghizadeh2020}, and photophysical properties of point defects \cite{QPOD, IMP2D, Ali2023}. However, some databases contain both TMD- and hBN-based quantum emitters and use the same functional for all materials \cite{C2DB,QPOD,Ali2023}. This results in an overall comprehensive database but not all functionals might be equally good for all materials. In contrast, there are also exclusive hBN defects databases \cite{QPOD,Ali2023}, however, there do not contain the complete optical fingerprint of many hBN defects.\\
\indent In this work, we develop a searchable online-database specific to fluorescent defects in hBN based on DFT with the highly accurate HSE06 functional by Heyd, Scuseria, and Ernzerhof \cite{10.1063/1.1564060}, which predicts properties in good agreement with the experiment \cite{Sajid2020,Reimers2018,Cholsuk2023}. To identify color centers of hBN, Ref. \cite{Cholsuk2023} has demonstrated that it should not rely on a single property because this likely leads to defect misassignment from the aforementioned property of various defect localizations. This work consequently collects the hBN database with comprehensive properties, namely electronic structure, zero-phonon line (ZPL), excitation and emission dipole orientations, lifetime, radiative transition rate, and spin multiplicity among 158 defects, including vacancies, substitutional defects, antisites, as well as defect complexes (combinations of them). Some defect structures are shown in Fig.~\ref{fig:workflow}.\\
\indent Having considered the hBN defects, they can be additionally classified according to the total spin of the system, comprising singlet, doublet, and triplet states. We focus especially on the triplet and singlet states as their electronic structures can be applied not only as a single photon source but also for quantum sensing \cite{Stern2022,Mu2022,Gottscholl2020,Zhou2023} and quantum memory applications \cite{Nateeboon2024,Cholsuk2024}. For quantum sensing, a recent example is the negatively charged boron vacancy. Its intrinsic triplet state allows it to have the splitting at $m_s$ = 0 and $\pm 1$, which supports the optically detected magnetic resonance (ODMR) measurement \cite{Stern2022,Zhou2023,Mu2022}. Meanwhile, relaxation can also exist through the non-radiative transition via the intersystem-crossing pathway; hence, the singlet electronic structure also needs to be unraveled. For a quantum memory, the triplet configuration supports the $\Lambda$ structure for writing and retrieving processes as has been proposed recently to implement in hBN \cite{Cholsuk2024,Nateeboon2024}. As such, in this database, if the neutral-charge defects are doublet states, the singly positive and negative charge states are applied to make them triplets.\\
\indent Therefore, our comprehensive database of color centers in hBN serves as another comprehensive resource. Complementing with the browsing website interface via \url{https://h-bn.info}, one can search for potential defect candidates that contribute to certain properties. Also, the incorporation of optimized lattice structures facilitates further theoretical investigations.

\section{Methods} \label{sec:method}
\subsection{Structural optimization for ground and excited states}
The Vienna Ab initio Simulation Package (VASP) \cite{vasp1,vasp2} with the projector augmented wave (PAW) method for the pseudopotentials \cite{paw,paw2} was performed to investigate all hBN structures. While the ground-state configuration can be obtained by default Gaussian smearing, the excited-state one needs manual electron occupation from the so-called $\Delta$SCF method \cite{Jones1989}. With this constraint method, both spin-up and spin-down transitions were considered. While the hBN monolayer was treated by adding the vacuum layer with 15 \AA~separated from the neighboring repeated cell to exclude layer-layer interaction, most experiments are performed using bulk structures. This difference may cause the inconsistency between simulation and experiment in the aspect of ionization energies, charge stabilities, and charge transition level \cite{badrtdinov_dielectric_2023,amblard_universal_2022,smart_fundamental_2018,wang_layer_2020}. However, the ZPL energy from a defect-defect transition is altered only slightly \cite{krecmarova_extrinsic_2021,turiansky_boron_2021}. Quantitatively, in some defects, the redshift of ZPL by 0.3 eV from monolayers can be expected \cite{winter_photoluminescent_2021}. All defects were in turn added at the center of 7$\times$7$\times$1 supercell as it had been verified not to have neighboring-cell effects. All defect geometries were relaxed by fixing the cell until the total energy and all forces were converged at 10$^{-4}$ eV and 10$^{-2}$ eV/\AA, respectively. Spin configurations; singlets, doublets, and triplets, were constrained by the difference in the number of unpaired electrons controlled by NUPDOWN in VASP. Last, the underestimation of the hBN bandgap was resolved by the HSE06 functional with the single $\Gamma$-point scheme.

\subsection{Zero-phonon-line calculation}
A single dominant peak observed in the experimental photoluminescence is related to the zero-phonon line (ZPL). This ZPL is defined as the transition without phonons involved. As such, this can be computed from the total energy difference between the ground and excited configurations. It is worth noting that while the zero-point energy difference between ground and excited states is typically around 1 meV to 0.1 eV, depending on a defect \cite{benedek_symmetric_2023, mishra_giant_2019}, our calculations assume that the zero-point energy is canceled out (only the energy difference between ground and excited state configurations matters). As noted earlier, the spin pathway between spin up and down is considered separately to conserve the radiative transition; this results in two ZPLs for certain defects if both spin transitions exist.

\subsection{Excitation and emission polarizations}
Excitation and emission polarizations can be extracted from the transition dipole. In principle, the polarizations were orthogonal to their dipoles; hence, both dipoles were rotated by 90$^\circ$ and taken modulo 60$^\circ$ to obtain the nearest angle to the crystal axis. Although the hexagonal lattice is spaced 120$^\circ$, after 180$^\circ$, the angle from the crystal axis will be identical. To make the theoretical polarization compatible with polarization-resolved PL experiments, both dipoles were projected onto the $xy$-plane to compute the in-plane polarization visibility.\\
\indent In order to calculate the dipoles, the wavefunctions after structural optimization can be extracted using the PyVaspwfc Python code \cite{pyWave}. This also implies that the wavefunctions between ground states and excited states are not necessarily the same. This leads to two types of dipoles; excitation and emission dipoles. The former is responsible for the transition from the wavefunction of the most stable ground states to the wavefunction of the excited state without geometry relaxation. Meanwhile, the latter describes the transition from the wavefunction of the most stable excited state to the wavefunction of the most stable ground state. This relation can be  expressed as
\begin{equation}
\boldsymbol{\mu} = \frac{i\hbar}{(E_{f} - E_{i})m}\bra{\psi_{f}}\textbf{p}\ket{\psi_{i}},
\label{eq:dipole}
\end{equation}
where the initial and final wavefunctions denoted by $\psi_{i/f}$ and the respective eigenvalues of the initial/final orbitals are indicated by $E_{i/f}$. Electron mass is denoted by $m$, and a momentum operator is denoted by $\mathbf{p}$. As the wavefunctions are taken from different structures, the modified version of PyVaspwfc is needed instead \cite{Davidsson2020}. Since the dipoles can contribute both in-plan and out-of-plane directions, such contribution can be differentiated by considering $\mu_z$ in the following expression $\boldsymbol{\mu} = \abs{\mu_x}\hat{x} + \abs{\mu_y}\hat{y} + \abs{\mu_z}\hat{z}$. If it is equal to 0, the dipole becomes purely in-plane (with respect to the $xy$-plane of hBN crystal). This can be quantified by the so-called {\it{linear in-plane polarization visibility}}.

\subsection{Radiative transition rate and lifetime}
Radiative transition is the transition between two defect states that conserves spin polarization. This can be quantified by the following equation
\begin{equation}
\Gamma_{\mathrm{R}}=\frac{n_D e^2}{3 \pi \epsilon_0 \hbar^4 c^3} E_0^3 \mu_{\mathrm{e}-\mathrm{h}}^2,
\label{eq:dipole}
\end{equation}
where $\Gamma_{\mathrm{R}}$ is the radiative transition rate. $e$ is the electron charge; $\epsilon_0$ is vacuum permittivity; $E_0$ is the ZPL energy; $n_{\mathrm{D}}$ is set to 1.85, which is the refractive index of the host hBN in the visible \cite{Vogl2019}; and $\mu_{\mathrm{e}-\mathrm{h}}^2$ is the modulus square of the dipole moment computed by Eq.\ \ref{eq:dipole}. Note that the hBN refractive index can be varied across the samples. Finally, the lifetime is calculated by taking the inverse of the transition rate. It should be noted that the calculated lifetime can be different from the experimental value due to the Purcell effect \cite{Vogl2019}. That is, most experiments attach hBN layers to a substrate, which can alter the density of states that the emitter can couple to. The dipole emission pattern and the emitter lifetime are therefore affected.

\subsection{Non-radiative transition rate and lifetime}
Despite the occurrence of radiative transition, there are instances where transitions between defects occur through intersystem crossing. This transition is so-called non-radiative. The rate of this transition can be obtained by the following equations. 
\begin{equation}
\Gamma_{\mathrm{NR}}=\frac{2 \pi}{\hbar} g \sum_{n, m} p_{i n}\left|\left\langle f m\left|H^{\mathrm{e}-\mathrm{ph}}\right| i n\right\rangle\right|^2 \delta\left(E_{f m}-E_{i n}\right),
\label{eq:fermi-golden}
\end{equation}
where $\Gamma_{\mathrm{NR}}$ is the non-radiative transition rate between electron state $i$ in phonon state $n$ and electron state $f$ in phonon state $m$. $g$ is the degeneracy factor. $p_{\text{in}}$ is the thermal probability distribution of state $|i n\rangle$ based on the Boltzmann distribution. $H^{\mathrm{e}-\mathrm{ph}}$ is the electron-phonon coupling Hamiltonian. Finally, the lifetime is the inverse of the transition rate, similar to the radiative case.

\subsection{Quantum efficiency}
Once the radiative and non-radiative transition rates can be computed, the quantum efficiency can be then acquired from
\begin{equation}
    \eta = \frac{\Gamma_{\mathrm{R}}}{\Gamma_{\mathrm{R}}+\Gamma_{\mathrm{NR}}}. \label{eq:q_eff}
\end{equation}

\subsection{Photoluminescence}
Photoluminescence $L(\hbar\omega)$ can be computed from
\begin{equation}
    L(\hbar\omega) = C\omega^3 A(\hbar\omega),
\end{equation}
where $C$ is a normalization constant from fitting experimental data, and $A(\hbar\omega)$ is the optical spectral function given by
\begin{equation}
    A(E_{ZPL} - \hbar\omega) = \frac{1}{2\pi}\int_{-\infty}^{\infty}G(t)\exp(-i\omega t-\gamma|t|) dt,
\end{equation}
where $G(t)$ is the generating function of $G(t) = \exp(S(t) - S(0))$, and $\gamma$ is a fitting parameter. Then the time-dependent spectral function $S(t)$ is attained by
\begin{equation}
    S(t) = \int_0^\infty S(\hbar\omega)\exp(-i\omega t)d(\hbar\omega),
\end{equation}
where $S(\hbar\omega)$ is a \textit{total} Huang-Rhys (HR) factor, which can be calculated from 
\begin{equation}
    S(\hbar\omega) = \sum_k s_k\delta(\hbar\omega - \hbar\omega_k),
\end{equation}
where $s_k$ is the \textit{partial} HR factor for each phonon mode $k$, given by
\begin{equation}
    s_k = \frac{\omega_k q_k^2}{2\hbar},
\end{equation}
where $q_k$ is the configuration coordinate, provided by the following expression
\begin{equation}
    q_k = \sum_{\alpha,i}\sqrt{m_\alpha}\left(R_{e,\alpha i} - R_{g,\alpha i}\right)\Delta r_{k,\alpha i}.
\end{equation}
$\alpha$ and $i$ run over the atomic species and the spatial coordinates $(x,y,z)$, respectively. $m_{\alpha}$ is the mass of atom species $\alpha$. $R_g$ and $R_e$ are the stable atomic positions in the ground and excited states, respectively, while $\Delta r_{k,\alpha i}$ is the displacement vector of atom $\alpha$ at the phonon mode $k$ between the ground and excited states.
\subsection{Quantum memory properties}
All quantum memory properties have been evaluated based on the off-resonant Raman protocol, which relies on the dynamics in the $\Lambda$ electronic structures. For more details on the computational techniques and the data production of the quantum memory properties, we refer the interested reader to Refs.~\cite{Nateeboon2024,Cholsuk2024}.

\section{Database Acquisition}
Fig.~\ref{fig:workflow} depicts the collection of this database. In the beginning, 158 impurities covering groups III-VI and their complexes are created in the hBN monolayer. Their total spins are then determined by spin multiplicity: singlet, doublet, and triplet. As listed in Tab.~\ref{table:defect_list}, we found 95 out of 158 defects acting as triplet/singlet states under the neutral-charge state, while the rest preferred the doublet state. For these doublet defects, they will be charged with $\pm 1$ charges. This guarantees that every defect can have a triplet and singlet configurations. The structures of both total spins are, in turn, optimized, and this can be treated as a ground state. For an excited state, the electrons are manually occupied using the $\Delta$SCF approach. Taking into account the electronic transitions separately, for instance, from spin up to up or from spin down to down, this yields 257 triplet electronic structures for ground states, 257 triplet electronic structures for excited states, and 211 singlet electronic structures for ground states. We note that as of now, only triplet states are considered for their excited states. Other spin configurations will be added later to the database. As a consequence, the properties can now be extracted by the methodology described in Sec.~\ref{sec:method}.

\begin{figure*}[ht!]
    \centering
    \includegraphics[width = 0.8\textwidth]{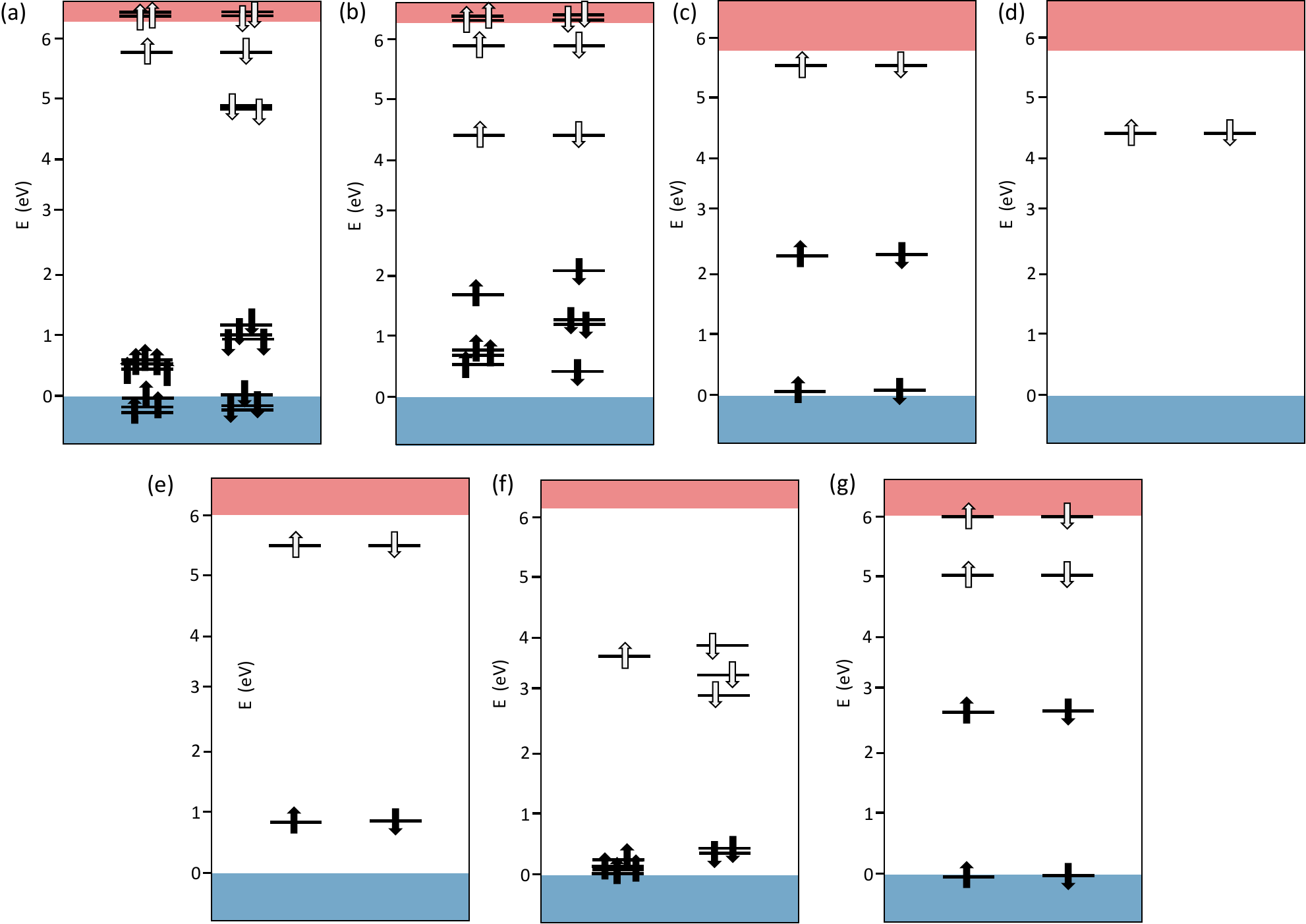}
    \caption{Kohn-Sham electronic structure of (a) triplet V$_\text{B}^{-1}$, (b) singlet V$_\text{B}^{-1}$, (c) singlet-state C$_\text{N}^{-}$, (d) singlet-state C$_\text{B}^{+}$, (e) singlet-state C$_\text{B}$C$_\text{N}$, (f) triplet-state C$_\text{N}$V$_\text{B}$, and (g) singlet-state C$_\text{B}$V$_\text{N}$. The blue and red regions represent valence and conduction bands, respectively. The filled and unfilled arrows indicate the occupied and unoccupied states, whereas their directions correspond to spins up and down.}
    \label{fig:band}
\end{figure*}
\section{Results and Discussion}
In this section, the initial part involves comparing the predicted photophysical properties within our database with those obtained from other DFT calculations and experimental data. Subsequently, we illustrate the distributions of ZPL, radiative lifetime, and polarization misalignment for each periodic table.  Lastly, we provide a detailed explanation of how to use the database effectively.
\subsection{Database benchmarking}
As defects in this database cover triplet and singlet spin multiplicities, our benchmarking can, in turn, cover both triplet and singlet states. Figs.~\ref{fig:band}(a) and \ref{fig:band}(b) display the electronic transitions of the negative boron vacancy (V$_\text{B}^{-1}$) for triplet and singlet states, respectively. This is because a neutral boron vacancy was found to act as the double state, as listed in Tab.~\ref{table:defect_list}. This defect leads to positive and negative charge triplet defects after single charging. We found that V$_\text{B}^{-1}$ triplet state supports only one possible transition pathway, which is from spin down to spin down, as the unoccupied defect state in the spin-up pathway is localized too close to the conduction band. This is in excellent agreement with other DFT calculations \cite{Chen2021,Ivady2020}. For this spin-down transition, the ZPL yields 2.08 eV, which is slightly higher than the DFT value by 0.37 eV \cite{Ivady2020} and the experimental value by 0.46 eV \cite{Gottscholl2020}. This can be justified by the difference in used functionals.\\
\indent Despite such little ZPL overestimation of V$_\text{B}^{-1}$, we have also benchmarked our results with other defects, including C$_\text{N}^{-}$, C$_\text{B}^{+}$, C$_\text{B}$C$_\text{N}$, C$_\text{N}$V$_\text{B}$, and C$_\text{B}$V$_\text{N}$. We found that all electronic transitions agree well with other DFT calculations \cite{Abdi2018,Jara2020, Chen2021,Sajid2018}. In addition, our calculated ZPL for C$_\text{B}$V$_\text{N}$ is 1.89 eV, close to other DFT reports at 1.95 eV \cite{Sajid2018}. Note that the ZPLs for the rest of the defects were not given. Thus, we can deduce that the ZPL is sensitive to a defect type and a functional.\\
\indent While the limitation of ZPL has been shown, other properties collected in this database can increase the rate of accuracy in defect identification in comparison with experiments. This has been explicitly demonstrated by our prior work through the example of C$_2$C$_2$ and C$_2$C$_\text{N}$ defects \cite{Cholsuk2023}. In this database, we extended the comprehensive properties (rather than ZPL) to cover 257 triplet transitions. Consequently, we believe that this database has the potential to serve as a valuable resource for designing and conducting experiments, offering essential insights and guidance.
\begin{figure*}[ht]
    \centering
    \includegraphics[width = 0.925\textwidth]{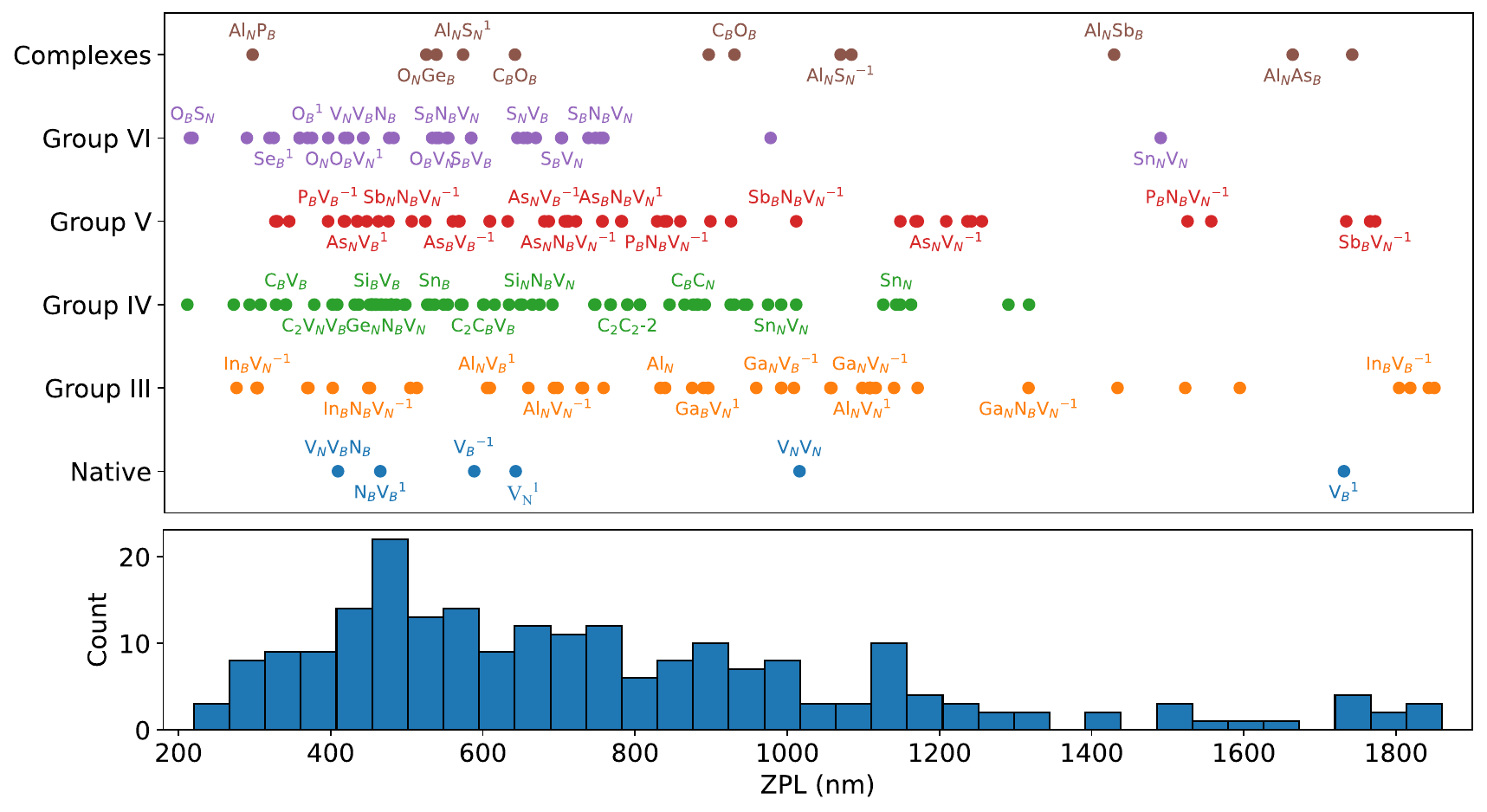}
    \caption{Distribution of ZPLs classified by the periodic table together with histogram. Note that the annotation next to each data point is shown only for certain defects. All of them can be found in the database.}
    \label{fig:distribution_zpl}
\end{figure*}

\begin{figure*}[ht]
    \centering
    \includegraphics[width = 0.925\textwidth]{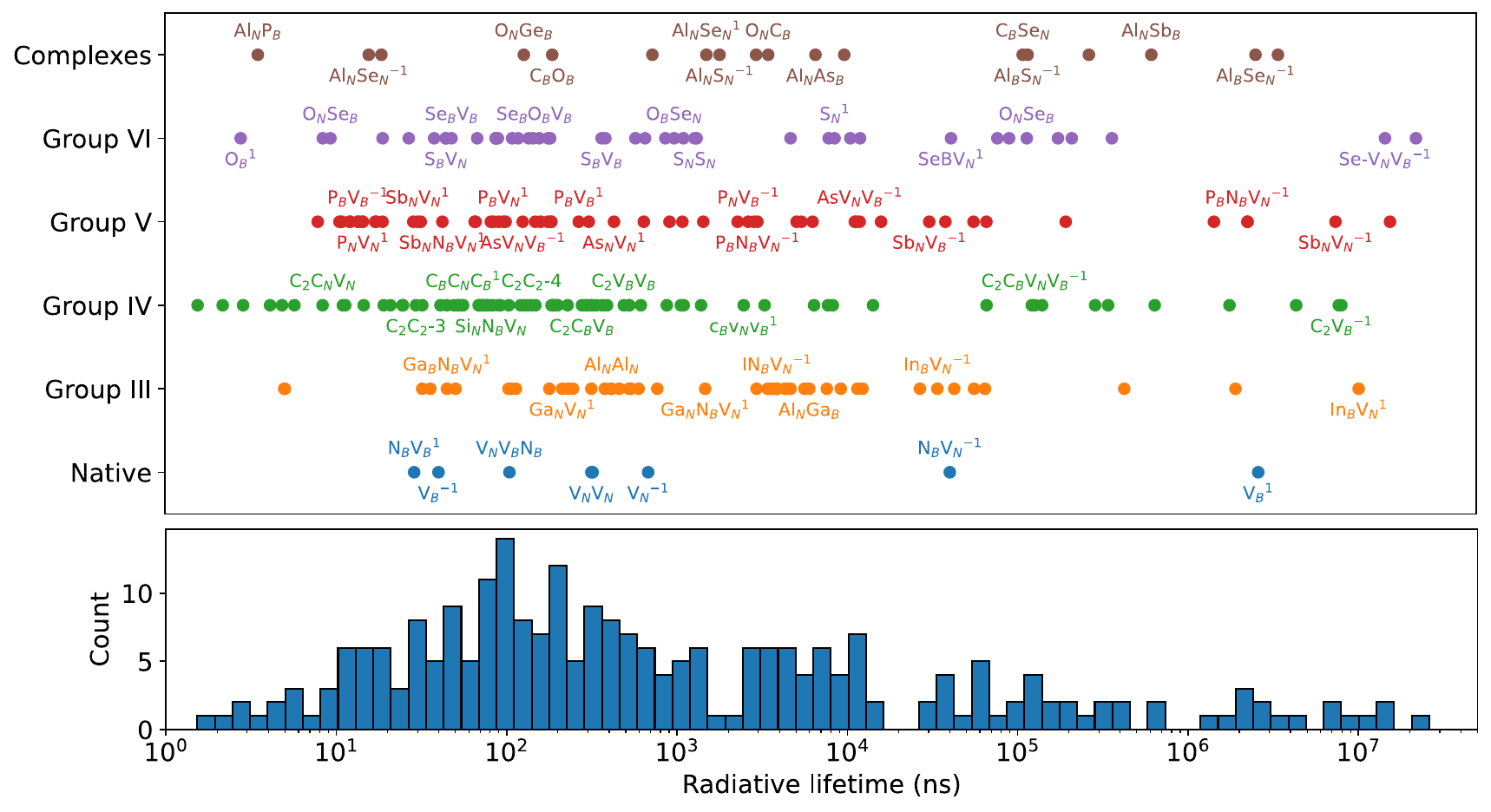}
    \caption{Distribution of radiative lifetime classified by the periodic table together with histogram. Note that the annotation next to each data point is shown only for certain defects. All of them can be found in the database.}
    \label{fig:distribution_lifetime}
\end{figure*}

\begin{figure*}[ht]
    \centering
    \includegraphics[width = 0.925\textwidth]{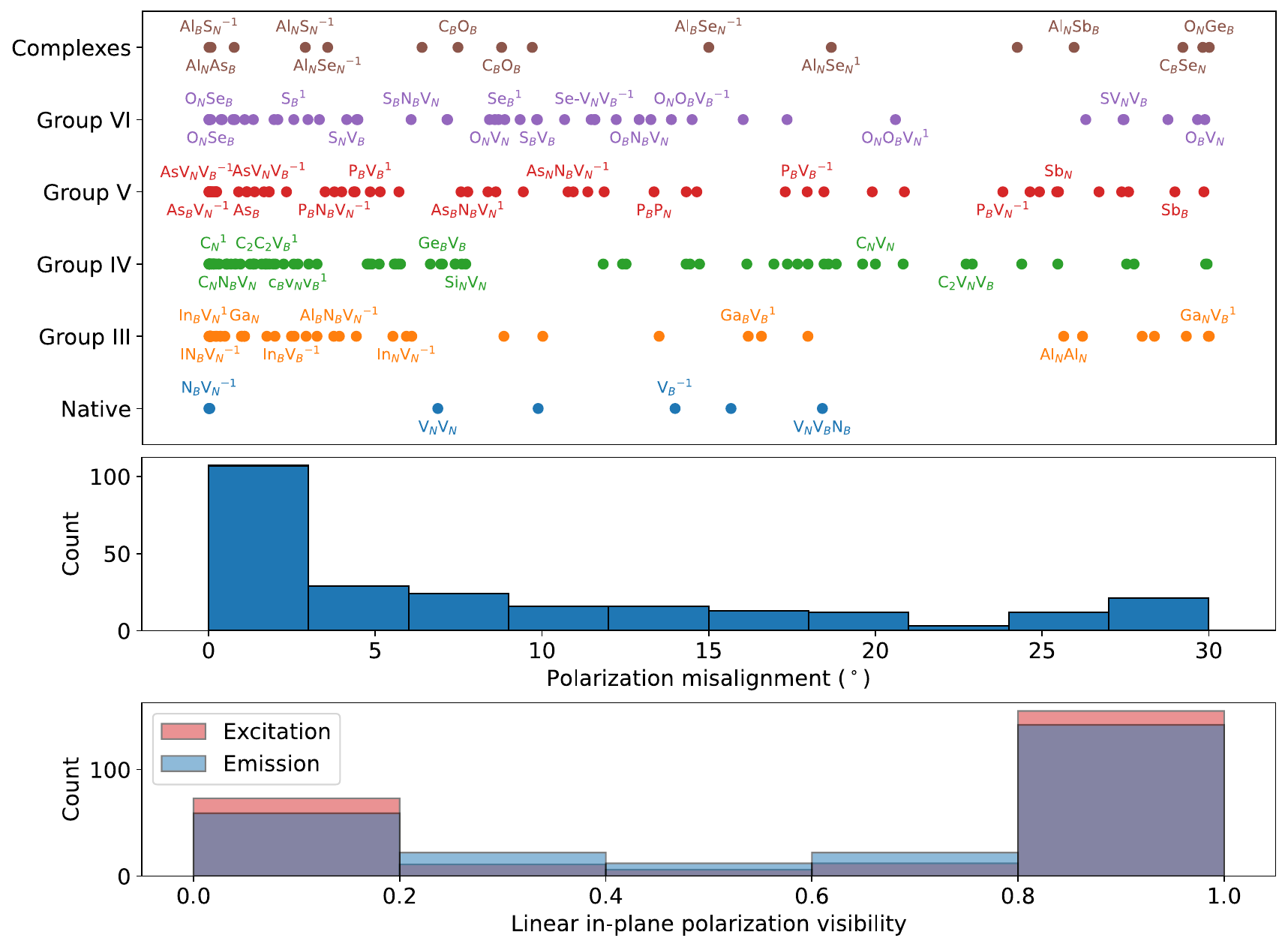}
    \caption{Distribution of polarization misalignment classified by the periodic table together with histogram. Polarization misalignment is obtained from the difference of angles between excitation and emission polarizations. Both polarization angles are listed in the database.  Note that the annotation next to each data point is shown only for certain defects. All of them can be found in the database.}
    \label{fig:distribution_misalignment}
\end{figure*}
\subsection{Database screening}
After individually charging the doublet defects and considering the ZPL for each feasible transition pathway, 257 triplet electronic transitions are characterized as depicted in Fig.~\ref{fig:distribution_zpl}. We found that most defects belong to the ZPLs around the visible range, especially from 400 nm to 600 nm. This can be justified by the two-level system localization. That is, as the pristine bandgap of hBN is around 6 eV, the occupied and unoccupied defect states are mostly localized in the deep-lying region, which is around 1-2 eV and 4-5 eV, respectively. As a consequence, the transition energy between these two defect states yields in the range of 2 to 3 eV, which corresponds to the ZPLs around 400-600 nm. Taking into account the relationship between ZPLs and the periodic table, both are uncorrelated, as can be seen from the unpredictable tendency of each group. In fact, this is reasonable as the periodic table may only affect the number of defect states, not the state location. This consequently agrees well with the transition energy distribution revealed previously \cite{Cholsuk2022}. This result can then be deduced that the ZPL cannot be simply estimated from the periodic table but needs the actual calculation.\\
\indent Turning to consider the radiative lifetime, this essentially corresponds to the ZPL pathway, as the lifetime can account for the transition rate between two defect states. Fig.~\ref{fig:distribution_lifetime} indicates the lifetime values categorized by each periodic group. We found that the lifetimes are dominant in the order of 30 $\mu$s to 1 ns, which is consistent with a typical range found in experiments \cite{Gottscholl2020,Kumar2023,PhysRevApplied.19.044011}. Comparing the distribution of ZPLs with that of radiative lifetime, they are uncorrelated. This is in agreement with another theoretical analysis, which found a lifetime relationship with the exciton energy, not the ZPL \cite{Gao2021}. Further, the relationship between the lifetime and the periodic table itself is also not found. This is reasonable because the transition rate, as shown in Eq.~\ref{eq:dipole}, is obtained from the optimized wavefunctions between ground and excited states, both of which strongly depend on the type of defects.\\
\indent Lastly, the polarization misalignment between excitation and emission polarizations is analyzed as illustrated in Fig.~\ref{fig:distribution_misalignment}.
It suggests that the majority of both polarizations are parallel among each other, leading to a tiny misalignment. For the group distribution, as the polarization relates to the transition dipole moment, it makes sense not to have a relationship between the periodic groups. In addition to the parallel polarizations, their majority is pointing out in an in-plane direction, as observed from the linear in-plane polarization visibility. This implies that most hBN defects inherit the polarized light. Here, we emphasize that the in-plane \textit{excitation} polarization is not necessarily required to result in the in-plane \textit{emission} polarization. This can be quantified by the linear in-plane polarization visibility between both polarizations. This behavior can be explained by the differences in wavefunctions, which are optimized differently for each polarization type. This phenomenon has also been observed in a recent experimental work of Ref.~\cite{Kumar2024}.

\section{Defect identification procedure}
The database offers multiple options for practical exploration of defect properties. For effective defect identification, it is important to recognize the limitations of DFT to ensure promising defects are not inadvertently overlooked. Accordingly, the step-by-step procedure is described below.\\
\indent One underlying experimental observation is the zero phonon line (the most prominent peak from the photoluminescence spectrum). This property is directly accessible in the database; however, it is recommended to furnish a range rather than a specific value. This entails conducting a search within a range of  $\pm$0.4 eV from the observed ZPL. The finite accuracy of DFT and residual strain from the experiment justify this range, rendering it incompatible with the unstrained defects documented in the database.\\
\indent While the ZPL range serves as a reliable indicator, additional insights can be obtained from the fabrication technique as well. For instance, if a defect is fabricated using ion implantation, we can infer certain defects based on the ions employed. 
Meanwhile, defects formed through the use of a scanning electron microscope are inclined to be carbon-related defects emerging as a result. \\
\indent Then, we can further narrow down the potential defect candidates by considering other photophysical properties such as lifetime, polarization orientation, etc. This way will enhance the accuracy of defect identification, as demonstrated earlier \cite{Cholsuk2024}. The database will be frequently updated with new defects and new defect properties. In addition, users can also request specific data (defects or properties) which will be prioritized for the next update. All data, including crystallographic data, can be downloaded in various forms.

\section{Conclusion and outlook}
This work proposes a database using a suitable DFT methodology to reveal ground and excited photophysical properties for triplet color centers in the hBN. Over 257 triplets and 211 singlets allow us to determine the relationship with the periodic table, which we found uncorrelated. Meanwhile, the property distribution suggests the dominant ranges of ZPLs, lifetime, and polarizations, corresponding to the experiments.\\
\indent Considering the quantum applications, all studied defects in this database inherit triplet electronic structures, which are in principle suitable for ODMR and the $\Lambda$ structure for quantum memory. However, the former still requires zero-field splitting and ODMR simulations for confirmation, while the latter has been added to this database through the performance metrics developed by our prior work \cite{Cholsuk2024}. As a consequence, this database can also be employed to pinpoint potential candidates suitable for quantum applications.\\
\indent In addition to being comprehensive for guiding an experiment, the interface of the open-source website has the potential to enable users to browse defect candidates with specific inherited properties. Furthermore, the database also provides optimized lattice structures that can directly perform DFT calculations. This database will be beneficial not only for experiments but also for theory in assigning microscopic origins.\\
\indent Outstanding additions to this database include the following: defect formation energy is required for analyzing which charged defects are indeed thermodynamically favorable. While we acknowledge this fact, the primary focus of this database is to present the properties of hBN triplet defects under the assumption of their experimental viability. The addition of defect formation energy is scheduled for the upcoming update (the database will be maintained and frequently updated with more defects and more defect properties). The other important parameters to be added are excited properties, i.e., vertical excitation energy, photoluminescence, and non-radiative transition, which are also planned for the next database update. Last but not least, as the hBN refractive index can be varied across several experiments, featuring this refractive index as an open parameter is scheduled to update so that one can obtain the corresponding transition rate and lifetime according to the specified refractive index.

\section*{Data availability}
All raw data from this work is available from the authors upon reasonable request. The database with browsing interfaces is freely accessible at \url{https://h-bn.info}.

\section*{Notes}
The authors declare no competing financial interest.

\begin{acknowledgments}
This research is part of the Munich Quantum Valley, which is supported by the Bavarian state government with funds from the Hightech Agenda Bayern Plus. This work was funded by the Deutsche Forschungsgemeinschaft (DFG, German Research Foundation) - Projektnummer 445275953 and under Germany's Excellence Strategy - EXC-2111 - 390814868. T.V. is funded by the Federal Ministry of Education and Research (BMBF) under grant number 13N16292. The authors acknowledge support by the German Space Agency DLR with funds provided by the Federal Ministry for Economic Affairs and Climate Action BMWK under grant number 50WM2165 (QUICK3) and 50RP2200 (QuVeKS). C.C. is grateful to the Development and Promotion of Science and Technology Talents Project (DPST) scholarship by the Royal Thai Government. The computational experiments were supported by resources of the Friedrich Schiller University Jena supported in part by DFG grants INST 275/334-1 FUGG and INST 275/363-1 FUGG. The authors gratefully acknowledge the Gauss Centre for Supercomputing e.V.\ (www.gauss-centre.eu) for funding this project by providing computing time on the GCS Supercomputer SuperMUC-NG at Leibniz Supercomputing Centre (www.lrz.de). The authors are grateful to Joel Davidsson for the source code of transition dipole moments for two wavefunctions.
\end{acknowledgments}
\newpage
\appendix
\onecolumngrid
\setcounter{table}{0}
\setcounter{figure}{0}
\section{Classification of hBN color centers}
Color centers in hBN are classified according to their total spins: single, doublet, and triplet. As this version of the database focuses on the triplet-state defects, the doublet defects are charged with positive and negative charges. The overall list of defects is shown in Tab.~\ref{table:defect_list}. 

\begin{table*}[h!]
\caption{Classification of spin multiplicity in neutral-charge hBN color centers. Note that C$_2$ stands for C$_\text{B}$C$_\text{N}$. C$_2$C$_2$-number refers to different lattice configurations, all of which are visualized in the database.}
    \centering
    \begin{tabular}{|l|l|l|l|l|l|l|}
   \hline
        \multicolumn{7}{|c|}{hBN color centers} \\ \hline
        \multicolumn{4}{|c|}{Triplet/singlet} & \multicolumn{3}{c|}{Doublet} \\ \hline
        Al$_\text{N}$ & C$_\text{B}$C$_\text{N}$C$_\text{N}$C$_\text{N}$ & O$_\text{B}$V$_\text{B}$ & Se$_\text{B}$V$_\text{B}$ & Al$_\text{B}$N$_\text{B}$V$_\text{N}$ & C$_\text{N}$ & S$_\text{B}$ \\ 
        Al$_\text{N}$Al$_\text{B}$ & C$_\text{B}$N$_\text{B}$V$_\text{N}$ & O$_\text{B}$V$_\text{N}$ & Se$_\text{N}$ & Al$_\text{B}$Se$_\text{N}$ & Ga$_\text{B}$N$_\text{B}$V$_\text{N}$ & Sb$_\text{B}$N$_\text{B}$V$_\text{N}$ \\ 
        Al$_\text{N}$Al$_\text{N}$ & C$_\text{B}$O$_\text{B}$ & O$_\text{N}$C$_\text{B}$ & Se$_\text{N}$V$_\text{B}$ & Al$_\text{B}$S$_\text{N}$ & Ga$_\text{B}$V$_\text{N}$ & Sb$_\text{B}$V$_\text{N}$ \\ 
        Al$_\text{N}$As$_\text{B}$ & C$_\text{B}$O$_\text{N}$ & O$_\text{N}$Ge$_\text{B}$ & Se$_\text{N}$V$_\text{N}$ & Al$_\text{B}$V$_\text{N}$ & Ga$_\text{N}$N$_\text{B}$V$_\text{N}$ & Sb$_\text{N}$N$_\text{B}$V$_\text{N}$ \\ 
        Al$_\text{N}$Ga$_\text{B}$ & C$_\text{B}$Se$_\text{N}$ & O$_\text{N}$N$_\text{B}$V$_\text{N}$ & Si$_\text{B}$ & Al$_\text{N}$N$_\text{$_\text{B}$}$V$_\text{N}$ & Ga$_\text{N}$V$_\text{B}$ & Sb$_\text{N}$V$_\text{B}$ \\ 
        Al$_\text{N}$P$_\text{B}$ & C$_\text{B}$V$_\text{B}$ & O$_\text{N}$O$_\text{B}$ & Si$_\text{B}$V$_\text{B}$ & Al$_\text{N}$Se$_\text{N}$ & Ga$_\text{N}$V$_\text{N}$ & Sb$_\text{N}$V$_\text{N}$ \\ 
        Al$_\text{N}$Sb$_\text{B}$ & C$_\text{B}$V$_\text{N}$ & O$_\text{N}$S$_\text{B}$ & Si$_\text{N}$ & Al$_\text{N}$S$_\text{N}$ & Ge$_\text{B}$ & SbV$_\text{N}$V$_\text{B}$ \\ 
        As$_\text{B}$ & C$_\text{N}$C$_\text{B}$C$_\text{B}$C$_\text{B}$ & O$_\text{N}$Se$_\text{B}$ & Si$_\text{N}$N$_\text{B}$V$_\text{N}$ & Al$_\text{N}$V$_\text{B}$ & In$_\text{B}$N$_\text{B}$V$_\text{N}$ & Se-V$_\text{N}$V$_\text{B}$ \\ 
        As$_\text{N}$ & C$_\text{N}$N$_\text{B}$V$_\text{N}$ & O$_\text{N}$Se$_\text{N}$ & Si$_\text{N}$Si$_\text{B}$ & Al$_\text{N}$V$_\text{N}$ & In$_\text{B}$V$_\text{B}$ & Si-V$_\text{N}$V$_\text{B}$ \\ 
        C-V$_\text{N}$V$_\text{B}$ & C$_\text{N}$O$_\text{N}$ & O$_\text{N}$S$_\text{N}$ & Si$_\text{N}$V$_\text{B}$ & As$_\text{B}$V$_\text{B}$ & In$_\text{B}$V$_\text{N}$ & Si$_\text{B}$V$_\text{N}$ \\ 
        C$_\text{2}$C$_\text{2}$-1 & C$_\text{N}$V$_\text{B}$ & O$_\text{N}$V$_\text{B}$ & Si$_\text{N}$V$_\text{N}$ & As$_\text{B}$V$_\text{N}$ & In$_\text{N}$N$_\text{B}$V$_\text{N}$ & Si$_\text{N}$V$_\text{N}$ \\ 
        C$_\text{2}$C$_\text{2}$-2 & C$_\text{N}$V$_\text{N}$ & O$_\text{N}$V$_\text{N}$ & S$_\text{N}$ & As$_\text{N}$N$_\text{B}$V$_\text{N}$ & In$_\text{N}$V$_\text{B}$ & S$_\text{N}$-V$_\text{N}$V$_\text{B}$ \\ 
        C$_\text{2}$C$_\text{2}$-3 & Ga$_\text{N}$ & P$_\text{B}$ & Sn$_\text{B}$ & As$_\text{N}$V$_\text{B}$ & In$_\text{N}$V$_\text{N}$ & Sn$_\text{B}$V$_\text{N}$ \\ 
        C$_\text{2}$C$_\text{2}$-3 & Ge-V$_\text{N}$V$_\text{B}$ & P$_\text{B}$P$_\text{N}$ & Sn$_\text{B}$V$_\text{B}$ & As$_\text{N}$V$_\text{N}$ & N$_\text{B}$VN & V$_\text{B}$ \\ 
        C$_\text{2}$C$_\text{2}$-4 & Ge$_\text{B}$N$_\text{B}$V$_\text{N}$ & P$_\text{N}$ & Sn$_\text{N}$ & AsV$_\text{N}$V$_\text{B}$ & O$_\text{B}$ & V$_\text{N}$ \\ 
        C$_\text{2}$C$_\text{2}$-5 & Ge$_\text{B}$V$_\text{B}$ & P$_\text{N}$As$_\text{B}$ & S$_\text{N}$N$_\text{B}$V$_\text{N}$ & C$_\text{2}$C$_\text{2}$V$_\text{B}$ & O$_\text{N}$ & ~ \\ 
        C$_\text{2}$C$_\text{B}$V$_\text{B}$ & Ge$_\text{B}$V$_\text{N}$ & S-V$_\text{N}$V$_\text{B}$ & Sn$_\text{N}$V$_\text{B}$ & C$_\text{2}$C$_\text{B}$V$_\text{N}$V$_\text{B}$ & O$_\text{N}$O$_\text{B}$V$_\text{B}$ & ~ \\ 
        C$_\text{2}$C$_\text{B}$V$_\text{N}$ & Ge$_\text{N}$N$_\text{B}$V$_\text{N}$ & Sb$_\text{B}$ & Sn$_\text{N}$V$_\text{N}$ & C$_\text{2}$C$_\text{N}$ & O$_\text{N}$O$_\text{B}$V$_\text{N}$ & ~ \\ 
        C$_\text{2}$C$_\text{N}$V$_\text{B}$ & Ge$_\text{N}$V$_\text{B}$ & Sb$_\text{N}$ & S$_\text{N}$S$_\text{N}$ & C$_\text{2}$C$_\text{N}$V$_\text{N}$V$_\text{B}$ & P$_\text{B}$N$_\text{B}$V$_\text{N}$ & ~ \\ 
        C$_\text{2}$C$_\text{N}$V$_\text{N}$ & Ge$_\text{N}$V$_\text{N}$ & S$_\text{B}$N$_\text{B}$V$_\text{N}$ & S$_\text{N}$V$_\text{B}$ & C$_\text{2}$V$_\text{B}$ & P$_\text{B}$V$_\text{B}$ & ~ \\ 
        C$_\text{2}$V$_\text{B}$V$_\text{B}$ & In$_\text{N}$ & S$_\text{B}$V$_\text{B}$ & V$_\text{N}$ & C$_\text{2}$V$_\text{N}$ & P$_\text{B}$V$_\text{N}$ & ~ \\ 
        C$_\text{2}$V$_\text{N}$V$_\text{B}$ & O$_\text{B}$N$_\text{B}$V$_\text{N}$ & S$_\text{B}$V$_\text{N}$ & V$_\text{N}$V$_\text{B}$N$_\text{B}$ & C$_\text{B}$ & P$_\text{N}$N$_\text{B}$V$_\text{N}$ & ~ \\ 
        C$_\text{2}$V$_\text{N}$V$_\text{N}$ & O$_\text{B}$Se$_\text{N}$ & Se$_\text{B}$ & V$_\text{N}$V$_\text{N}$ & C$_\text{B}$C$_\text{N}$C$_\text{B}$ & P$_\text{N}$V$_\text{B}$ & ~ \\ 
        C$_\text{B}$C$_\text{N}$ & O$_\text{B}$S$_\text{N}$ & Se$_\text{B}$N$_\text{B}$V$_\text{N}$ & ~ & C$_\text{B}$V$_\text{N}$V$_\text{B}$ & P$_\text{N}$V$_\text{N}$ & ~ \\ \hline
  \end{tabular}
    \label{table:defect_list}
\end{table*}

\twocolumngrid
\bibliography{main}
\end{document}